\documentclass[%
 reprint,
 amsmath,amssymb,
 aps,
]{revtex4-2}

\usepackage{graphicx}
\usepackage{dcolumn}
\usepackage{bm}
\usepackage{xcolor}
\usepackage{url}
\usepackage{subfig}

\newcommand{\megnet}{\textsc{megnet} }
\newcommand{\settset}{\textsc{set2set} }
\newcommand{\matbench}{\textsc{matbench} }

\graphicspath{{./plots/}, {./figures/}}

\begin{document}

\preprint{APS/123-QED}

\title{Entropy-based Active Learning of Graph Neural Network Surrogate Models for Materials Properties}

\author{Johannes Allotey}
\affiliation{%
School of Physics, University of Bristol, BS8 1TL, UK\
 }
\author{Keith T. Butler}%
 \email{keith.butler@stfc.ac.uk}
\affiliation{%
 Scientific Machine Learning Research Group, Scientific Computing Department, Rutherford Appleton Laboratory, Science and Technology Facilities Council, Didcot, OX11~0DQ, UK\\
}

\author{Jeyan Thiyagalingam}
\affiliation{%
Scientific Machine Learning Research Group, Scientific Computing Department, Rutherford Appleton Laboratory, Science and Technology Facilities Council,  Didcot, OX11~0DQ, UK\\
}


\begin{abstract}
Graph neural networks, trained on experimental or calculated data are becoming an 
increasingly important tool in computational materials science. Networks, once trained,
are able to make highly accurate predictions at a fraction of the cost of experiments 
or first-principles calculations of comparable accuracy. However these networks 
typically rely on large databases of labelled experiments to train the model. In 
scenarios where data is scarce or expensive to obtain this can be prohibitive. By building a 
neural network that provides a confidence on the predicted properties, we are able
to develop an active learning scheme that can reduce the amount of labelled data 
required, by identifying the areas of chemical space where the model is most uncertain. We present a scheme for coupling a graph neural network with a Gaussian process to 
featurise solid-state materials and predict properties \textit{including} a measure
of confidence in the prediction. We then demonstrate that this scheme can be used in
an active learning context to speed up the training of the model, by selecting the 
optimal next experiment for obtaining a data label. Our active learning scheme can double the rate at which the performance of the model on a test data set improves with additional data compared to choosing the next sample at random. This type of uncertainty quantification and active learning has the potential to open up new areas of materials science, where data are scarce and expensive to obtain, to the transformative power of graph neural networks.
\end{abstract}

\maketitle


\section{Introduction}

Machine learning (ML) has become an important tool in almost every modern scientific discipline, and materials' science is no exception \cite{butler2018machine}. In particular, the proliferation of data in recent years has given rise to the advent of deep learning approaches, where neural networks with hundreds of thousands or even millions of parameters learn to infer trends from relatively unstructured data \cite{chen2019graph, xie2018crystal, park2020developing, lee2021transfer, raza2020message}. In materials' science, network architectures based on graphs, Graph Neural Networks (GNNs), have proved to be successful both for molecules and condensed matter systems\cite{xie2018crystal, chen2019graph, raza2020message}. GNNs allow the encoding of domain knowledge about connections (bonds) into the topography of the underlying neural network architecture. Despite the great success and promise of GNNs for materials' science applications a number of questions remain open. In this paper we address two of these questions (i) how much can we trust the results of a GNN on a previously unseen sample, and (ii) how can we train a GNN if data is scarce or labelled data is expensive to obtain?

\begin{figure}[ht!]
\centering
\includegraphics[width=\linewidth]{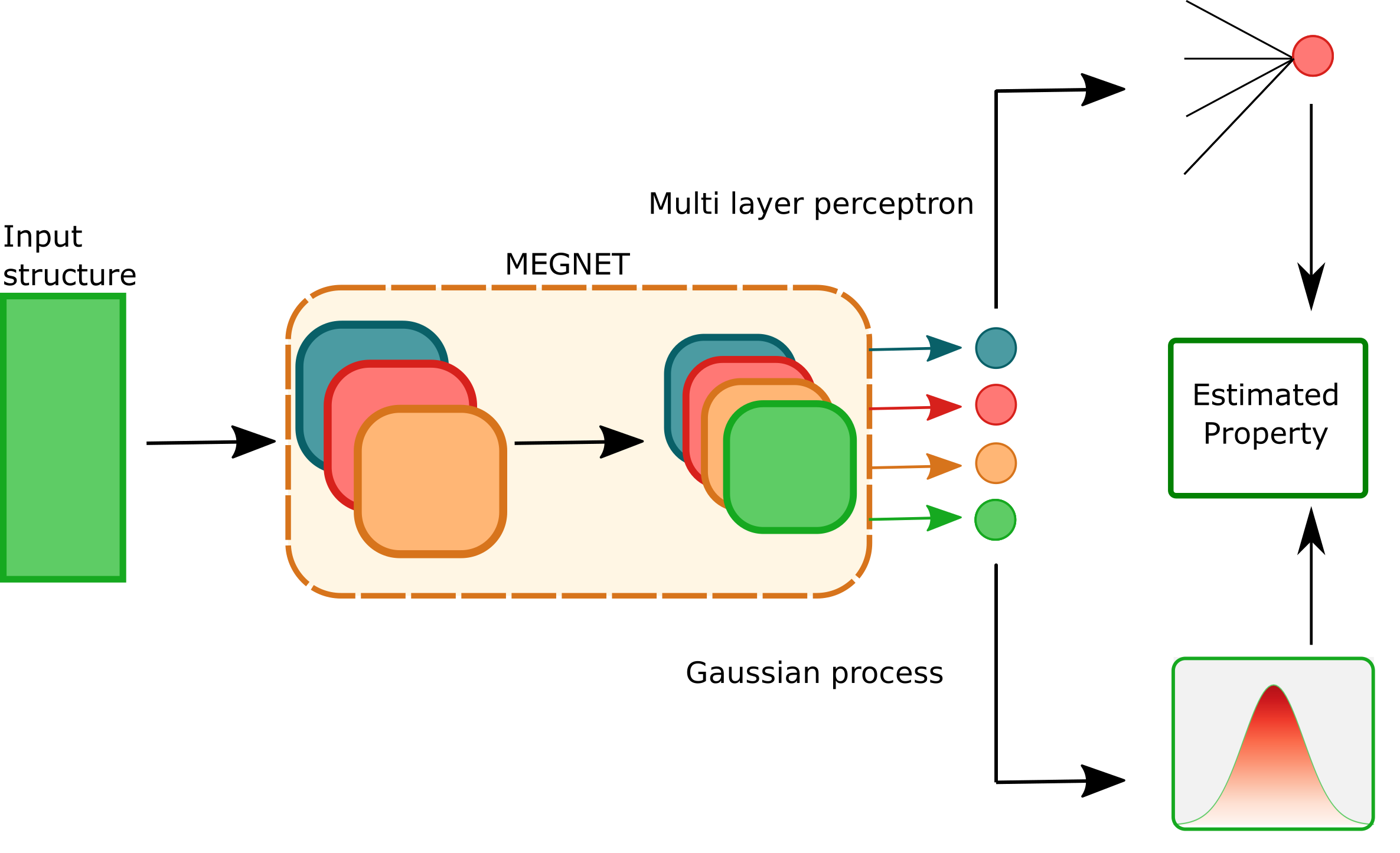}
\caption{The structure of the convolution fed Gaussian process (CFGP). The graph network part of \megnet learns representations for materials, in the standard model these are fed to a multi-layer perceptron for function approximation, in the CFGP these are used as input for a Gaussian process for function approximation.} 
\label{fig:latent_space}
\end{figure}

Uncertainty quantification (UQ) of machine learning models, and neural networks in particular, is  currently an area of intense research. Conventional neural networks return a single output value per property that they are trained on (classes or values in a classification or a regression scenario). These values contain neither information about how confident the network is, nor how the output varies with variation of the input. In a recent high-profile example, a network trained to classify COVID-19 infection from lung X-rays classified a cat as COVID-19 infected lung\cite{mallick2020sample}. In this case an estimation of certainty would have been very beneficial. 

Bayesian neural networks (BNNs) provide NNs with a principled UQ\cite{mackay1992practical}. However, BNNs are not scalable to most practical problems. Although various approximations, eg Monte Carlo dropout and mean field variational inference, \cite{blundell2015weight, gal2016dropout} improve scalability of BNNs, this comes  at the cost of UQ reliability. Deep ensembles provide generally good estimates of UQ but are not \textit{a priori} reliable. 

Alternative machine learning methods, rooted in Bayesian statistics, such as Gaussian processes do provide principled, and scalable UQ in many scenarios \cite{Bilionis2017}. GPs have proved very successful in materials modelling with a range of studies recently employing them for applications such as calculating energies and dynamics \cite{raimbault2019using, meyer2020geometry}. The major barrier in applying GPs to practical problems is the necessity to construct meaningful representations (or features) of the data as an input to the GP. Several elegant examples, such as the SOAP descriptor, have been proposed, but still need to be derived for each new system\cite{musil2021physics}.

One avenue to addressing the problem of feature engineering is to rely on representation learning or feature learning, an area of machine learning that focuses on automatically discovering the features that best describe the underlying (possibly hidden) characteristics of the data. Representation learning thus helps to not only better understand a dataset, but also to produce better outcomes. Although specialised architectures can be built to extract minimal representations of data \cite{tschannen2018recent}, even the hidden layers of simple neural networks progressively capture these representations \cite{bengio2013representation}.Representation learning has been instrumental in a number of areas\cite{gidaris2018unsupervised, kopf2021latent}.

In this work we combine a GNN with a GP by performing feature learning with the GNN and using the features learned by the GNN as input to a GP, thus circumventing the need for manual feature engineering. This approach has recently been demonstrated to work well for calculating adsorption energies for catalysis \cite{tran2020methods} and builds on a number of notable recent efforts towards uncertainty quantification for machine learning of molecular and materials properties\cite{janet2019quantitative, scalia2019evaluating}.  We demonstrate that the recently published \megnet architecture can be used to learn representations that can be applied in a GP to predict formation energy of a crystal. 

We first show that the latent space of the \megnet model is structured in such a way that information about the target properties can be inferred. We then demonstrate a well-calibrated GP with reliable UQ on the prediction of formation energies. We use this UQ to address the issue of using GNNs in scenarios where labelled data is scarce. We use the uncertainties calculated on previously unseen examples to choose the next training point for the ML model. This so-called entropy sampling-based active learning procedure is applied to training a model for formation energy prediction and is shown to significantly outperform random sampling of the next training point.  

We are confident that the methods and models presented in this paper can be applied to a range of properties and scenarios in materials' science, for example when training data is difficult to obtain such as dielectric properties, phonons or high quality electronic structures \cite{morita2020modelling}. Active learning has already demonstrated  transformative power in the field of molecular quantum chemistry\cite{smith2019approaching},  we hope to contribute towards its application in condensed matter materials.

\section{Latent space and predicted properties}

We first investigate whether a latent space representation of a material can be useful 
as an input vector for a Gaussian process. In order to do this we have extracted the outputs from the graph network section of \megnet. To be precise we use the values from the layer labelled ``Dense 32'' from \cite{chen2019graph}, this network was trained for 1,000 epochs on 1,800 randomly chosen materials from the full dataset, before being used to extract the activations for the same 1,800 materials. After a material is input to the network, we extract the values of the activations in this layer, this 32-dimensional vector serves as a compressed representation of the material. Visualisation of higher dimensional data is well-known to be difficult for human cognition, thus we reduce the dimensions of the latent space for visulaisation purposes, using the t-distributed stochastic neighbour embedding (t-SNE)\cite{hinton2002stochastic}. t-SNE is a non-linear and stochastic dimensionality reduction technique that projects the original data points (32-dimensions) to two-dimensional space.

The t-SNE-derived 2D distribution of the activation vectors for the dataset (see Dataset section for more details) are presented in Figure ~\ref{fig:latent_space}. In this figure we have coloured the points according to their formation energy. From this Figure, it is clear that there is a structure in the latent space relating to formation energy; points tend to have similar formation energies to those clustered close by. This gives us confidence that the graph network is learning a representation (or features) that contains the relevant information required for building a subsequent model for approximating the formation energy. 

\begin{figure}[ht!]
\centering
\includegraphics[width=\linewidth]{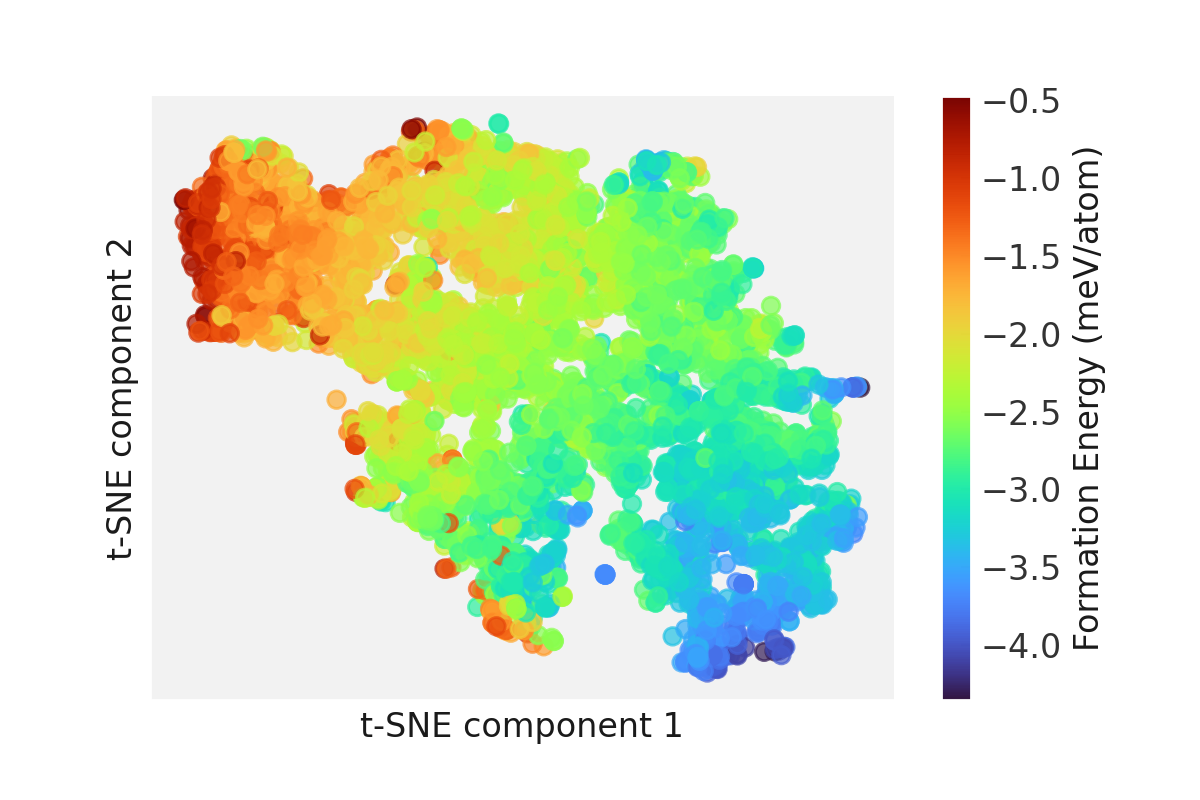}
\caption{A reduced dimensionality plot of the activations extracted from the first dense layer of the \megnet model. The activations are projected onto a 2D space using the t-SNE algorithm. The points are coloured according to the formation energy of the material.} 
\label{fig:latent_space}
\end{figure}

\section{A network-fed Gaussian process}

Having established that there is a structure of the latent space that is related to the output property of interest, we now develop a Gaussian process (GP) model to infer the relationship. Using the notation of Tran and co-workers\cite{tran2020methods} we refer to this as a convolution-fed Gaussian process (CFGP). The GP uses the 32-dimensional vector from the GNN as the input vector - full details of the GP are given in the Models section.

To characterise the similarity between samples, we use the Laplacian kernel
\begin{equation}
    k(x, y) = a^2\exp\left(\frac{-||x - y||}{l}\right) 
\end{equation}

where the similarity $k$ between two data samples at points $x$ and $y$ depends on a function of the difference between these points and an amplitude $a$ and length scale $l$ of the kernel. The full details of how values for $a$ and $l$ are obtained are given in the Methods section.

The CFGP is trained on 7,800 materials from the original dataset. Where 1,800 materials were previously used to train the \megnet model and the additional 6,000 materials are identified during an active learning procedure (see Methods for details). Note that during the active learning procedure the weights of the \megnet model are not updated, only the GP is trained, however at the end of the active learning we re-optimized \megnet and the GP on the new training set of 7,800 materials. We then test the model on 1,460 previously unseen materials to evaluate the performance.

\begin{figure}[ht!]
\centering
\includegraphics[width=\linewidth]{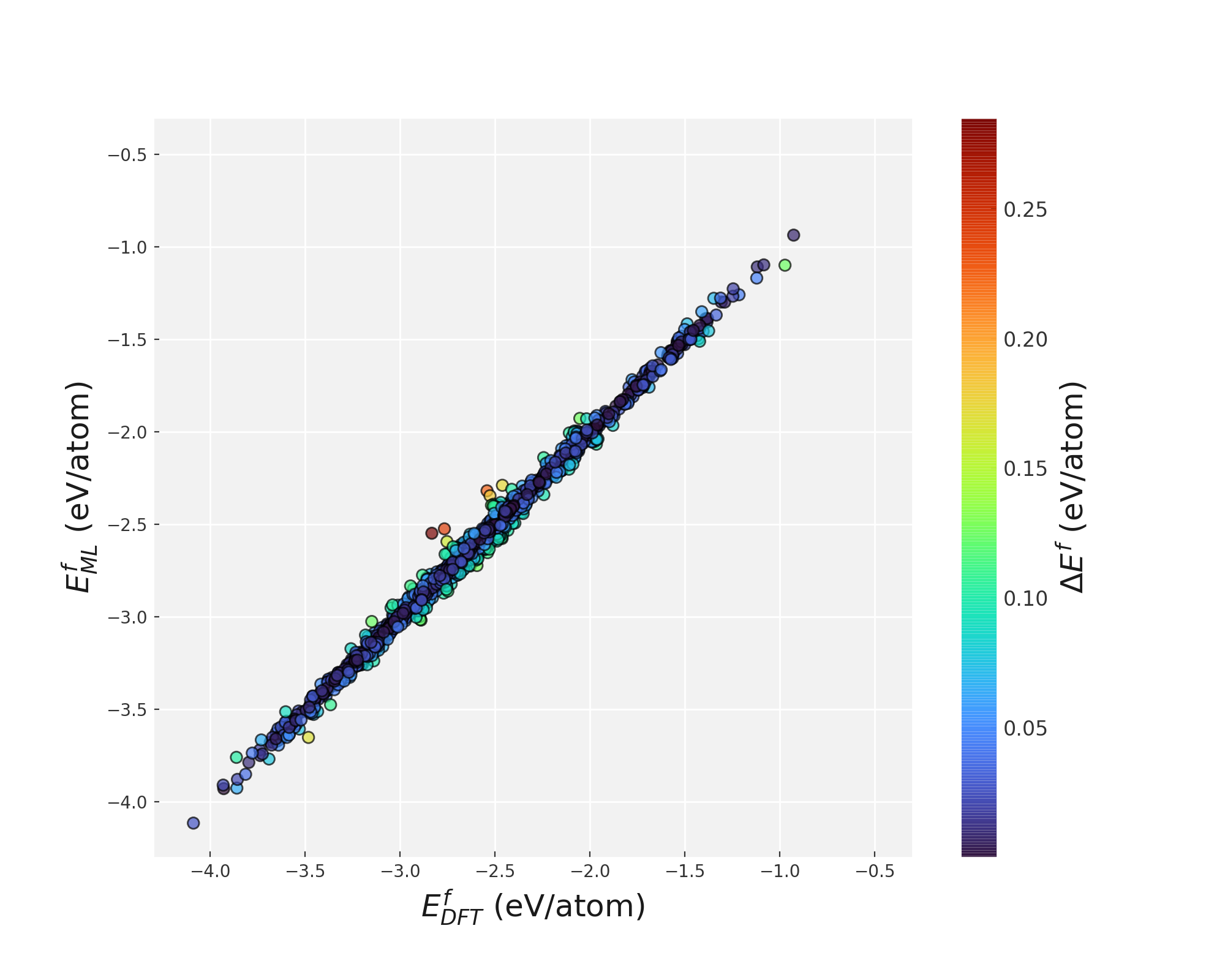}
\caption{The performance of the CFGP model for predicting formation energies on a test set of 1460 materials. The model has a mean absolute error of 0.0277 meV/atom, with an $r^2$ value of 0.99.} 
\label{fig:ml-performance}
\end{figure}

Figure \ref{fig:ml-performance} plots the CFGP predicted formation energy against the DFT calculated formation energy, for a test set of 1,460 materials not used to train the model. We can observe a high degree of correlation between the predicted and ground-truth (DFT) values. The test set data has a mean absolute error (MAE) of 0.0277 eV/atom, a mean squared error of 0.002 eV/atom. This performance is comparable to the performance of other GNN-based methods as reported on the \matbench benchmark dataset\cite{dunn2020benchmarking}, where \megnet had and MAE of 0.0417 eV/atom. We note that the training and test sets that we apply here are somewhat more restricted than the \matbench data, explaining our lower MAE.

This demonstrates that the CFGP can perform the task of predicting properties from materials structure with comparable fidelity to GNNs employing multi-layer perceptron function approximators. The advantage of the GP is that we also obtain UQ, or degree of confidence on the prediction. In the section that follows this, we evaluate the resulting UQ performance.

\section{Calibration of uncertainty quantification}

To assess the quality of the UQ element in our CFGP we compare the uncertainty estimates in the training set to the residuals between the predicted and true values. The work of Tran and \textit{et al.} \cite{tran2020methods} provides an excellent, lucid introduction to benchmarking UQ methods, outlining recently proposed protocols \cite{kuleshov2018accurate, levi2019evaluating}. We briefly outline the criteria for good UQ methods here but, highly recommend the aforementioned work of Tran \textit{et al.}\cite{tran2020methods}. 

A good uncertainty estimate should be well-calibrated, sharp, and disperse. In our context, well-calibrated means that the residuals generally fall within the range of the error estimates (in our case twice the standard deviation of the GP). Sharpness implies that the error estimates are not unnecessarily large; conservative large error-estimates give good calibration, but are not necessarily useful for assessing how the model will perform. Finally disperse means that the error-estimates cover a range of values, so that the model does not simply predict a uniform estimate for all samples.

\begin{figure}[ht!]
\centering
\includegraphics[width=\linewidth]{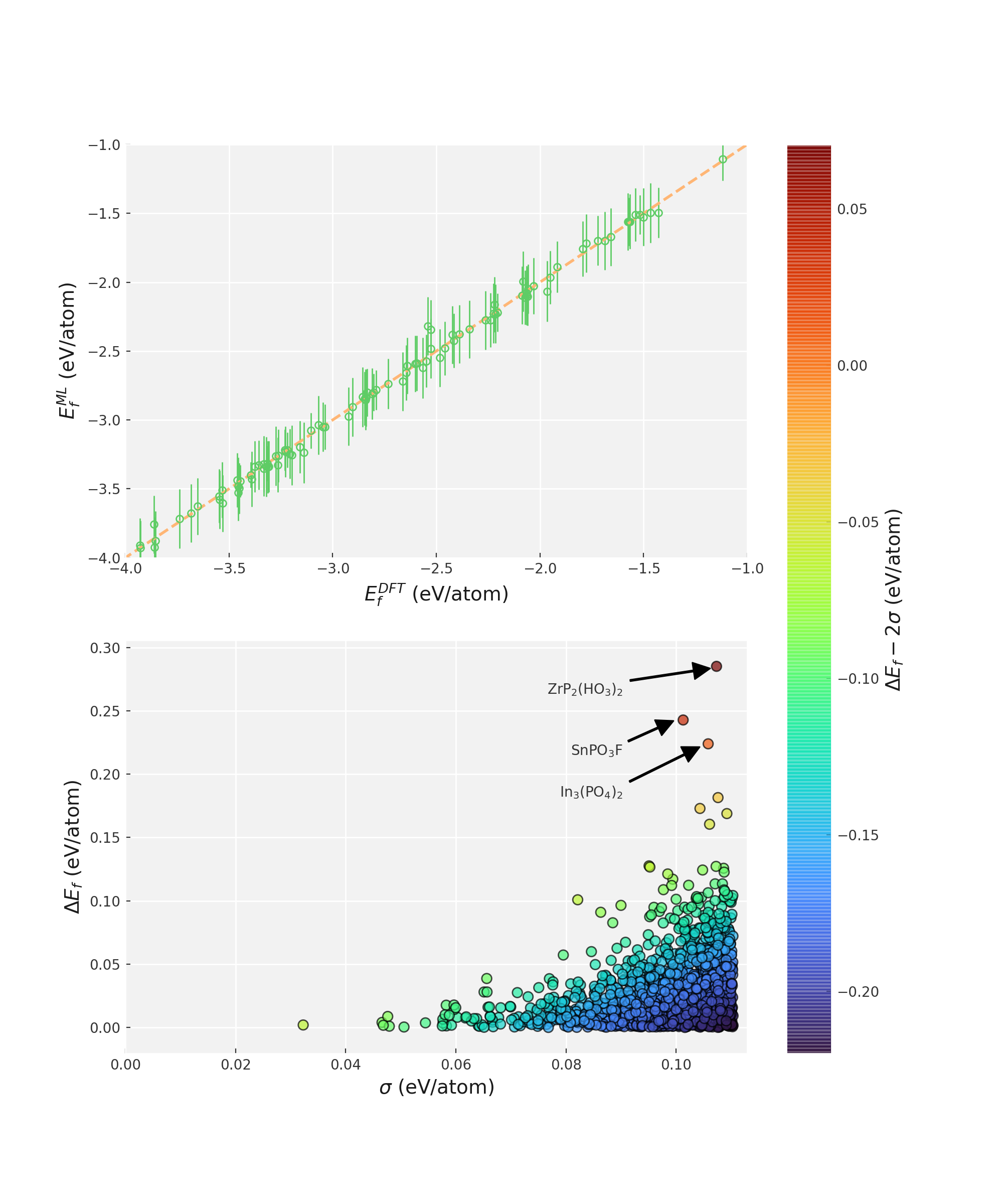}
\caption{The effectiveness of uncertainty quantification by the Gaussian process. Top plot shows 100 randomly selected points, the CFGP energy is plotted against the DFT energy, the error-bars are set to twice the standard deviation obtained from the CFGP. Bottom plot shows the residuals (difference between DFT energy and CFGP energy) versus the standard deviation obtained from the CFGP.} 
\label{fig:uq-model}
\end{figure}

In Figure ~\ref{fig:uq-model} we show the performance of the resulting UQ. To provide more clarity, the upper plot shows a randomly selected set of 100 samples from the test set. Specifically, the CFGP energies are plotted against the DFT energies with error bars equal to twice the standard deviation ($2\sigma$) of the GP; in a standard distribution, 95\% of the data should fall within $2\sigma$. We can observe that, within the sample set, the great majority of points fall within $2\sigma$ of the DFT energy. In the lower panel we plot the residuals against $\sigma$. We use the the same data as in Figure~\ref{fig:ml-performance} for 1,460 test materials, not used in training. This plot demonstrates that validation samples with larger residuals generally do have larger uncertainty estimates. We also colour the points by the difference between the residual and $2\sigma$, indicating which points fall outside the error estimates. 

 We note in Figure~\ref{fig:uq-model} that there are a few materials that stand out with particularly high errors and also with a large discrepancy between the UQ and the error in the prediction. Among these, three particularly poor predictions are labelled in the figure, we note that all three systems have phosphate motifs. The phosphate motif is relatively unusual in the context of an inorganic materials database. In addition two of the materials, ZrP$_2$(HO$_3$)$_2$ and  SnPO$_3$F have an apical oxygen in the phosphate replaced with P-H and P-F bonds respectively. These kinds of bonds are unusual and are therefore probably not present in large amounts in the training set. The scarcity of these motifs can explain why the model fails to predict them with good accuracy. We suggest that our UQ here is possibly over-cautious, but does generally identify well cases where the results of the model are likely to be unreliable. The calibration of the UQ could potentially be improved by exploring alternative kernels for the GP or using different layers of \megnet as input vectors, this will be the subject of further investigations, for now we assess how well the model can perform in an active learning procedure. 

\section{Active learning}

We now explore how the UQ demonstrated in the previous section can be further leveraged for active learning. In active learning, the ML model not only learns the relationships in the underlying data from labelled samples, but also chooses which unlabelled sample should be labelled next in order to best improve the performance and generality of the model. There are a number of techniques that can be used to select the points to be labelled next \cite{del2020assessing}. Recently an approach based on query by committee has been used to accelerate training for a quantum chemistry dataset \cite{smith2019approaching}.

In our case, we apply an entropy-based sampling approach, which chooses the unlabelled sample with the highest uncertainty and uses that as the next sample to be labelled. The uncertainty represents the entropy of the samples from an information theory perspective. Hence, opting to label the sample with greatest uncertainty next maximises the entropy reduction in the sampling process.  In the context of the GP, a point with high uncertainty is likely to be found in an area of feature space that is more sparsely labelled than other areas, thus choosing this point for labelling favours a more representative sampling of all regions of feature space. 

\begin{figure}[ht!]
\centering
\subfloat[][Schematic representation of our active learning process]{\includegraphics[width=0.9\linewidth]{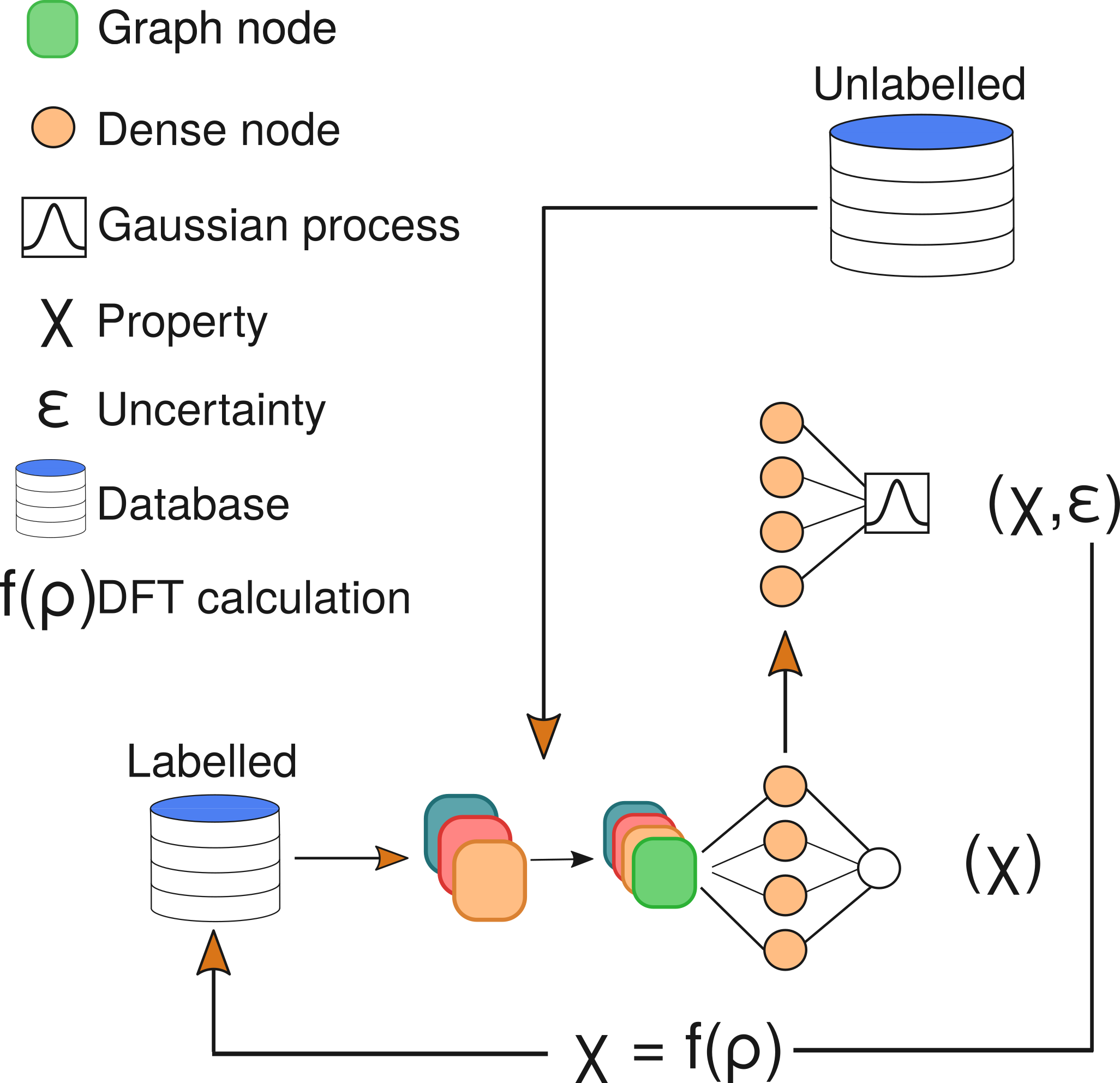}}\\
\subfloat[][Performance of our active learning process versus random sampling]{\includegraphics[width=0.9\linewidth]{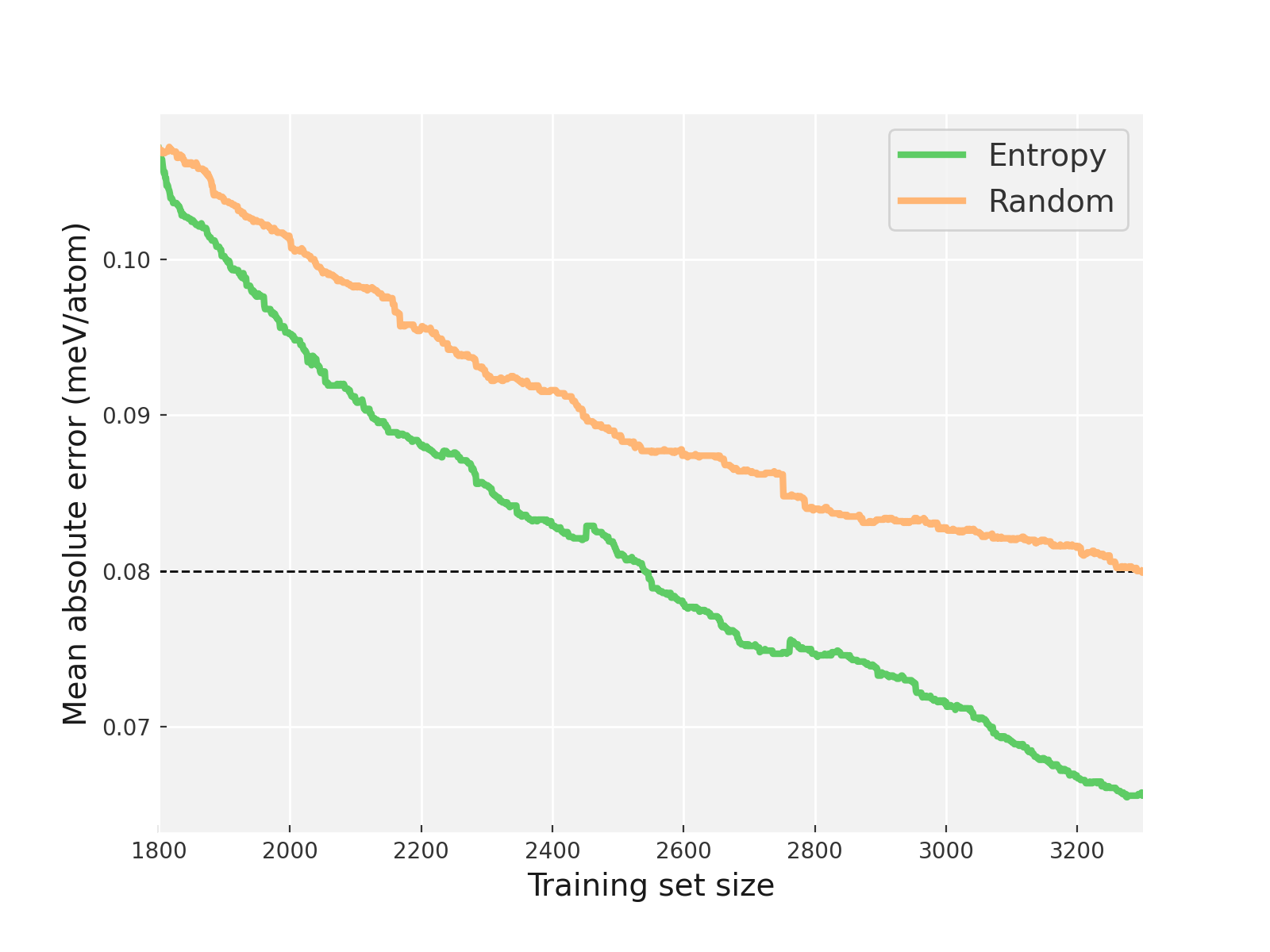}}
\caption{The process and effectiveness of entropy sampling based active learning. Upper; the active learning procedure, labelled data are passed through \megnet and activations at the dense layer extracted to train a Gaussian process, unlabelled data are then passed through the same process and predictions and uncertainties obtained, samples with the greatest uncertainty are then labelled and the procedure loops back around. Lower; the mean absolute error on the test set of the CFGP with different sampling methods. The entropy-based sampling method chooses the next sample to label based on the uncertainty in prediction of the unlabelled samples, the random sampling method chooses at random.} 
\label{fig:active-learning}
\end{figure}

A schematic illustration of our process is shown in the upper panel of Figure ~\ref{fig:active-learning}. First, a trained \megnet is passed the labelled data and the activations of the dense layer are used to feed a GP for training (a CFGP). Then the unlabelled data is passed through the CFGP providing predictions with uncertainties. Here, the sample with the greatest uncertainty is chosen to be labelled. Then the new larger labelled dataset is used to re-train the Gaussian process. The overall process continues until some predefined end point. In our case we started with 1,800 labelled materials from the dataset described in the Dataset section, we then run through 1,500 cycles of the active learning procedure evaluating the performance of the CFGP at each cycle model on all of the remaining materials whose labels were not used in training (i.e. the full set of materials minus the training set at that cycle).

The results of the entropy-based active learning process are presented in Figure~\ref{fig:active-learning} lower panel. For comparison we have also performed a sampling of new materials based on random selection from the unlabelled data.  We can see that the maximum entropy-based sampling approach significantly out-performs the random sampling-based approach. By the end of the 1,500 cycles of active learning, the model from the maximum entropy based approach outperforms the model from the random approach by $\sim$ 43\%. The enhancement in performance achieved by the random selection process after 1,500 additional samples is achieved by the entropy based sampling in fewer than half the samples; 740.

\section{Discussion}

We have demonstrated that Gaussian process trained on the outputs of a GNN, a CFGP, performs well for prediction, uncertainty quantification and active learning for the formation energy of materials. The CFGP model that we presented here has uses in a very wide range of materials science applications. In fact, a similar architecture was recently demonstrated to perform very well for prediction and uncertainty quantification in absorption energy prediction for catalysis\cite{tran2020methods}. This approach is particularly appealing as it combines the theoretical rigour of the Gaussian processes-based approaches, with the flexibility of the GNN approach. GNNs can, in principle, represent any materials or molecular system, with minimal user-defined input required, and therefore this approach should be easily adaptable to other domains of interest with minimal feature engineering .

The ability to place reliable confidence bounds on predicted properties is crucial for the application of ML methods and building trust in the outputs of models. For example, in the context of virtual high-throughput screening ML surrogate models are often applied as a filtering step \cite{davies2016computational, davies2019data, nyshadham2019machine, bobbitt2019molecular, li2017high}. Access to reliable UQ in this context is very attractive as it can be used in combination with threshold values to filter out prospective candidates. Reliable UQ can also reduce the number of false positives passing through the procedure or the number of false negatives being rejected, and thus could be used on-the-fly to assess edge cases at a higher level of fidelity. 

The potential application of active learning in materials and molecular science is being increasingly recognised \cite{lookman2019active, zhang2019bayesian}. In applications such as deriving new inter-atomic potentials\cite{podryabinkin2017active, smith2018less} or searching through materials space for optimal catalysts\cite{tran2018active}, the ability to optimise the choice of the next experiment is important. Active learning is particularly appealing in scenarios where data is scarce and expensive to obtain. For energy materials design properties such as high-quality electronic structure \cite{pilania2017multi}, dielectric constants\cite{petousis2017high, morita2020modelling, noda2020descriptors, tawfik2020predicting}, effective masses\cite{davies2019descriptors} and defect properties\cite{sharma2020machine, gu2019machine} are some high-profile examples where data-driven approaches are possible, but obtaining large sets of labelled data can be prohibitive to applying deep learning approaches. In these scenarios the availability of a method for reducing the number of samples required to train accurate, and reliable models is rather critical to the application of deep learning approaches in materials' science.

We note that there is definite room for improvement in the active learning strategy applied here. As demonstrated in the large errors on a family of phosphate materials, our approach may not be sampling the chemical space as effectively as possible. Improving the calibration of the UQ would be beneficial in active learning. There is scope for introducing chemical and statistical heuristics in order to improve the coverage. As an example, one could sample several new materials to label per active learning cycle, but ensure that each one comes from a distinct region of the latent space by enforcing a cut-off (possibly based on the kernel length of the GP). Additionally choice of kernel for the GP may affect how well calibrated the UQ is, this is an ongoing area of research. We also note that the active learning procedure was run for a further 4,500 cycles after the process described in the section Active Learning. However the advantage of using the entropy-based sampling diminishes as the process continues because the probability of randomly choosing useful material for training approaches that of actively choosing good materials by the entropy-based method as we sample larger portions of the available additional data points (in our case 7,195 materials). Thus, it is clear that the degree of advantage to be gained by applying the active learning approach will depend on the size of the non-labelled data and will likely increase as the unlabelled space increases.  However, our approach already demonstrates the true potential of active learning using an example where size of the unlabelled dataset is relatively small. With the number of unexplored inorganic materials estimated in the region of $10^{11}$ materials\cite{davies2016computational}, active learning will have an important role to play in navigating this space.

\section{Conclusions}

We have presented an active learning approach based on convolution-fed Gaussian processes, which is capable of greatly enhancing the learning rate of deep neural networks for predicting the properties of materials, when compared to random sampling. Our approach uses the fact that the latent space of a graph neural network model provides a compressed representation of a 3D crystal in a space with meaningful correlations to the composition-structure-property relationship which the GNN has been trained to reproduce. We use this compressed representation to feed a Gaussian process model, which provide property estimates and uncertainty quantification (UQ) on those estimates. We demonstrate that the estimates obtained are competitive with state-of-the-art GNNs and that the UQ obtained is well calibrated and sharp. We then take advantage of this UQ to develop an active learning procedure where the training dataset is augmented on-the-fly sampling from unlabelled data based on the UQ obtained from the model. We show that this active learning procedure results in a much more rapid improvement in model performance compared to sampling the unlabelled data randomly. The methods presented in this paper will help increasing the confidence in the output of ML models by providing reliable estimates of (un)certainty. Furthermore, the active learning procedure can be extremely useful for training models in scenarios where labelled data is difficult or expensive to obtain. We hope that our methods can help to accelerate the application of machine learning for materials design.

\section{Dataset}

The dataset for the formation energy models consists of 10455 oxide materials and their DFT calculated energies from the Materials Project database\cite{Jain2013} the exact dataset is available to download and can be used in association with our open repository of this code \cite{git_gpnet}.

The dataset is initially split into three parts - 1800 samples for training \megnet, 7,195 samples available for updating the model during active learning and 1,460 samples kept completely separate for model evaluation purposes.

The data used in this study are available from reference \cite{git_gpnet_data}.

\section{Models}

Our models in this paper are based on the \megnet GNN architecture \cite{chen2019graph}. \megnet consists of an embedding layer, followed by a flexible number of graph neural network units. The output from the graphs is then converted to a consistent sized vector using the \settset transformer\cite{vinyals2015order}. This vector is then fed into a multi-layer perceptron feed-forward neural network to approximate the property of interest. In the CFGP, the vector is fed into a GP for function approximation.

\begin{figure}[ht!]
\centering
\subfloat[][]{\includegraphics[width=0.9\linewidth]{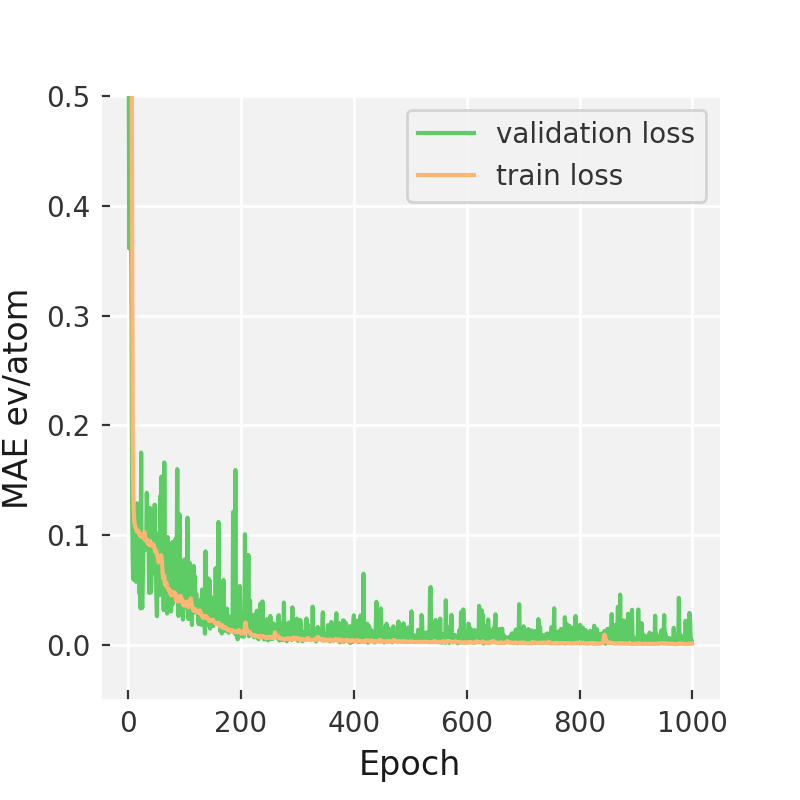}}\\
\caption{The training performance of the \megnet model. The mean absolute error on training and validation sets is plotted against epoch.} 
\label{fig:training}
\end{figure}

The \megnet model is initially trained on 1,800 randomly selected materials from the database. The model is trained for 1,000 epochs, until the validation error has equilibrated. The performance is shown in Figure~\ref{fig:training}.

Using the \megnet model trained as above, we build input vectors for the GP. Input vectors are obtained by passing an input to \megnet and extracting the activations from the ``Dense 32'' layer. The GP uses these vectors as input. The GP we employ uses the Laplacian kernel of Equation 1. We have implemented the GP using TensorFlow Probability \cite{dillon2017tensorflow}. The hyper-parameters $a$ and $l$ are obtained by optimising the negative log likelihood of GP on the training dataset of 1,800 materials. This process can be rather sensitive to the initial values of the hyper-parameters, starting from 1.0 for both $a$ and $l$, we obtain optimal values of 0.4281 and 2.3997 respectively. 

During active learning one new material is chosen at each cycle and added to the training set for the GP (note we do not re-train the \megnet model or the hyper-parameters of the GP at each step during the active learning). The GP is re-trained at each cycle of the active learning and the process was repeated for 6,000 cycles. At each cycle the performance of the GP on 1,460 independent samples, never used in any of the training steps is assessed.

For reference - the \megnet model took around 5 hrs to train; the hyper-parameter optimisation of the GP took 10,000 steps and 12 hrs and the active learning for 6,000 cycles took 7 hrs, all on a NVIDIA Quadro P5000 GPU. 

The code for the models, and trained model architectures are available at~\cite{git_gpnet} and~\cite{git_gpnet_data}, respectively.

\section*{Acknowledgements}

 This work was partially supported by Wave 1 of The UKRI Strategic Priorities Fund under the EPSRC Grant EP/T001569/1, particularly the “AI for Science” theme within that grant and The Alan Turing Institute. The ML models were trained using computing resources provided by STFC Scientific Computing Department's SCARF cluster and the PEARL cluster.

\section*{Data Access Statement}

All of the training data, trained neural networks and code for generating the training data for this study are openly available at \url{https://zenodo.org/record/4922828#.YMHksBIo-xI}.

A git repository containing the code used to build and train the neural networks, as well as notebooks to recreate them are available at \url{https://github.com/keeeto/gp-net}


\section*{Author Contributions}

KTB conceived, planned and steered the project. JA built, trained and applied the neural networks. KTB, JT and JA wrote the manuscript together. JT was involved in conception and establishing of the project and facilitated the work in the paper.

\bibliography{bibliography}

\end{document}